\documentclass[aps,twocolumn,showkeys,showpacs,preprintnumbers]{revtex4}
\usepackage{amssymb}
\usepackage{amsmath}
\usepackage{graphicx}

\begin{document}

\title{Generalized Gross-Pitaevskii equation adapted to the $U(5)\supset
SO(5)\supset SO(3)$ symmetry for spin-2 condensates}

\author{Y. Z. He$^1$, Y. M. Liu$^2$ and C. G. Bao$^{1,}$}
 \thanks{Corresponding author: stsbcg@mail.sysu.edu.cn}
 \affiliation{$^1$The State Key Laboratory of Optoelectronic Materials and Technologies, \\
School of Physics and Engineering, Sun Yat-Sen University,
Guangzhou, P. R. China}
 \affiliation{$^2$Shaoguan University, Shaoguan, 512005, P.R. China}



\begin{abstract}
A generalized Gross-Pitaevskii equation adapted to the
$U(5)\supset SO(5)\supset SO(3)$ symmetry has been derived and
solved for the spin-2 condensates. The spin-textile and the
degeneracy of the ground state (g.s.) together with the factors
affecting the stability of the g.s., such as the gap and the
level density in the neighborhood of the g.s., have been
studied. Based on a rigorous treatment of the spin-degrees of
freedom, the spin-textiles can be understood in a $N$-body
language. In addition to the ferro-, polar, and cyclic phases,
the g.s. might in a mixture of them when $0<M<2N$ ($M$ is the
total magnetization). The great difference in the stability and
degeneracy of the g.s. caused by varying $\varphi$ (which marks
the features of the interaction) and $M$ is notable.
 Since the root mean square radius $R_{\mathrm{rms}}$ is an observable,
efforts have been made to derive a set of formulae to relate
$R_{\mathrm{rms}}$ and $N$, $\omega$(frequency of the trap),
and $\varphi$. These formulae provide a way to check the
theories with experimental data.
\end{abstract}

\maketitle

\section{Introduction}

Since the success of trapping cold atoms via optical
trap,\cite{ste} the study of condensates of atoms with nonzero
spin is a hot topic. There are numerous literatures dedicated
to the study of the ground state (g.s.). For spin-2
condensates, the g.s. was found to have three phases, namely,
the ferro-phase, polar phase and cyclic phase, depending on the
features of the interaction.\cite{cvc,mko,mue} On the other
hand, it has been revealed that the total Hamiltonian of the
spin-2 systems is associated with the $U(5)\supset SO(5)\supset
SO(3)$ symmetry.\cite{pvi,ech} Since there are elegant theory
dealing with this symmetry, it is worthy to derive a
generalized Gross-Pitaevskii (GP) equation adapted to the
$U(5)\supset SO(5)\supset SO(3)$ symmetry. Since the solutions
of this equation could provide the details of the low-lying
states, we believe that the understanding to the spin-2
condensates could be thereby enriched.

We consider the case that the magnetic field $B=0$. The
symmetry-adapted GP equation is derived firstly. Then, by
solving the GP equation, both the spin-textiles and the spatial
wave functions of the low-lying states can be known. The
emphasis is placed on the g.s.. Since the stability of the g.s.
is affected by its environment, in addition to the g.s. itself,
special effort is made to study the environment, namely, the
width of the gap (the energy difference between the first
excited state and the g.s.) and the level-density in the
neighborhood of the g.s.. This is a topic less studied before.
Besides, the degeneracy of the g.s. is an important feature
which is also studied in this paper. The total magnetization
$M$ is considered as being conserved, which depends on how the
system is experimentally prepared. The influence of $M$ on the
spin-textiles and the degeneracy of the g.s. is found to be
great.

After we have obtained the spatial wave functions, physical
quantities related to the spatial degrees of freedom can be
known. In particular, we have calculated the root mean square
radius $R_{\mathrm{rms}}$. Since $R_{\mathrm{rms}}$ is an
observable, theoretical results and experimental data can be
compared with each other. This is a notable way to check the
theory. The GP equation will be firstly solved analytically
under the Thomas-Fermi approximation (TFA). A number of
formulae relating the physical quantities have been thereby
obtained, and the underlying physics can be understood in an
analytical way. Then, strict numerical calculations are
performed to check the validity of the TFA.

\section{Inherent symmetry and the Gross-Pitaevskii
equation}

A condensate with $N$ spin-2 neutral atoms trapped by an
isotropic harmonic potential $\frac{1}{2}m\omega^2 r^2$ is
considered, in which the spin-orbit coupling is assumed to be
weak. It is further assumed that the trap is not so weak (say,
$\omega\geq 100s^{-1}$) and the spin-dependent interaction is
weak so that the spin-modes are much lower than and separated
from the spatially excited modes. This is the basic assumption
of this paper. In other words, the following results hold only
for the systems adapted to this assumption. In this case a
group of excited states distinct in their spin-modes would
emerge in the neighborhood of the g.s.. In each of these states
(including the g.s.) all the particles have a common spatial
wave function which is most advantageous to binding and adapted
to a specific spin-textile. Thus one of these states with a
specific spin-mode $\gamma$ can be in general written as
\begin{equation}
 \Psi_{\gamma}
  =  \prod^N_{j=1}
     \phi_{\gamma}(\mathbf{r}_j)
     \Theta_{\gamma},
 \label{1}
\end{equation}
where $\phi_{\gamma}$ is the common spatial wave function and
$\Theta_{\gamma}$ is a total spin-state which is now unknown.
If the particles did not fall into the same state but have some
of them getting excited, the total energy would be thereby
higher. The group $\{\Psi_{\gamma }\}$ totally form a band
which is called the ground band. The states not in the ground
band will contain various spatially excited modes and will not
be studied in this paper.

When $\hbar\omega$ and $\lambda=\sqrt{\hbar/(m\omega)}$ are
used as units for energy and length (where $m$ is the mass and
is given at the one of $^{87}$Rb), the total Hamiltonian is
\begin{equation}
 H
  =  \sum_i
     h(i)
    +\sum_{i<j}
     V_{ij},
 \label{2}
\end{equation}
where $h(i)=-\frac{1}{2}\nabla_i^2+\frac{1}{2}r_i^2$. $V_{ij}
=\delta(\mathbf{r}_i-\mathbf{r}_j)\sum_s g_s P_s^{ij}$.
$P_s^{ij}=\sum_{m_s}|(\chi(i)\chi(j))_{s
m_s}\rangle\langle(\chi(i)\chi(j))_{s m_s}|$, $\chi(i)$ is the
spin-state of the $i$-th particle, and the two spins of $i$ and
$j$ are coupled to the combined spin $s$ and its z-component
$m_s$, $s=0$, 2, or 4. Obviously, $P_s^{ij}$ is the projector
onto the $s$-channel. $g_s$ is the strength related to the
scattering length of the $s$-channel.

We introduce an operator defined in the total spin-space as
$\tilde{\mathbf{V}}\equiv\sum_{i<j}\sum_s g_s P_s^{ij}$. It has
been proved that $\tilde{\mathbf{V}}$ is a linear combination
of a set of Casimir operators belonging to a chain of nested
algebra as $U(5)\supset SO(5)\supset SO(3)$.\cite{pvi,ech}
Consequently, the eigenstates and the eigenenergies of
$\tilde{\mathbf{V}}$ can be exactly known as
\begin{equation}
 \tilde{\mathbf{V}}
 \Theta _{\gamma}
  =  \tilde{E}_{vS}
     \Theta_{\gamma}.
 \label{3}
\end{equation}
Now, $\gamma$ represents a group of quantum numbers $v$,
$n_{\mathrm{tri}}$, $S$, and $M$, where $S$ and $M$ are the
total spin and its $Z$-component, $\nu$ and $n_{\mathrm{tri}}$
will be explained below, and $\Theta_{\gamma}\equiv\Theta_{v
n_{\mathrm{tri}}SM}$ is the related eigenstate. For
convenience, $M\geq 0$ is assumed.

When two spins are coupled to zero, they form a singlet pair
$\theta _{\mathrm{pair}}(ij)\equiv (\chi (i)\chi (j))_{0}$.
When three spins are coupled to zero, they form a triplet and
can be approximately denoted as
$\theta_{\mathrm{tri}}\approx(\chi\chi\chi)_0$ (this notation
is exact when $N\rightarrow\infty$).\cite{a1} In the triplet
every pair of spins are coupled to 2. It turns out that the
singlet pair and triplet are basic building blocks. An
eigenstate may contain $n_{\mathrm{pair}}$ singlet pairs and
$n_{\mathrm{tri}}$ triplets. The quantum number $v\equiv
N-2n_{\mathrm{pair}}$ is named the seniority, which is the
number of particles not in the singlet pairs. The number of
particles neither in the pairs nor the triplets is
$N_{\mathrm{free}}\equiv
N-2n_{\mathrm{pair}}-3n_{\mathrm{tri}}$, the total spin $S$ is
contributed by them. It has been proved that, when
$n_{\mathrm{pair}}$, $n_{\mathrm{tri}}$, $S$ and $M$ are given,
the totally-symmetric eigenstate $\Theta_{v
n_{\mathrm{tri}}SM}$ is unique. The four quantum numbers are
bound by the following conditions: (i) $N_{\mathrm{free}}\geq
0$, (ii) $N_{\mathrm{free}}\leq S\leq 2N_{\mathrm{free}}$,
(iii) $S=2N_{\mathrm{free}}-1$ is not allowed, and of course
(iv) $|M|\leq S$.\cite{pvi,agh}

Let us define $g_{04}=g_0-g_4$, $g_{24}=g_2-g_4$, and
$g_{(024)}=\frac{1}{3}(g_0+g_2+g_4)$. Then, the eigenenergy
associated with $\Theta_{\gamma}$ is \cite{pvi}
\begin{equation}
 \tilde{E}_{vS}
  =  a_1 N
    +a_2 N(N+4)
    +g_v v(v+3)
    +g_S S(S+1),
 \label{4}
\end{equation}
where $g_v=(-7g_{04}+10g_{24})/70$, $g_S=(-g_{24})/14$,
$a_1=\frac{11}{15}g_{04}-\frac{1}{42}g_{24}-\frac{5}{2}g_{(024)}$,
and
$a_2=-\frac{1}{15}g_{04}-\frac{1}{42}g_{24}+\frac{1}{2}g_{(024)}$.
Note that $\tilde{E}_{vS}$ does not depend on
$n_{\mathrm{tri}}$ implying that the levels might be degenerate
as shown below.\cite{agh}

Inserting Eq.(\ref{1}) into the Schr\"{o}dinger equation, we
have
\begin{equation}
 \langle
 \prod^N_{j=2}
 \phi_{\gamma}(\mathbf{r}_j)
 \Theta_{\gamma}|
 H-E_{\gamma}|
 \prod^N_{j=1}
 \phi_{\gamma}(\mathbf{r}_j)
 \Theta_{\gamma}
 \rangle
  =  0,
 \label{5}
\end{equation}
where the integration covers all the degrees of freedom except
$\mathbf{r}_1$. From this equation and making use of Eq.(3), it
is straight forward to obtain the symmetry-adapted GP equation
for the normalized $\phi_{\gamma}(\mathbf{r}_1)$ as
\begin{equation}
 [ h
  +\frac{2}{N}
   \tilde{E}_{vS}
   \phi_{\gamma}^*
   \phi_{\gamma}
  -\epsilon_{\gamma} ]
 \phi_{\gamma}
  =  0,
 \label{6}
\end{equation}
where $\epsilon_{\gamma}$ is the chemical potential. After
solving Eq.(6) we can obtain the total energy
$E_{\gamma}=N\epsilon_{\gamma}-V_{\mathrm{tot},\gamma}$, where
$V_{\mathrm{tot},\gamma}=\int|\phi_{\gamma}|^4 d\mathbf{r}\
\tilde{E}_{vS}$ is the total interaction energy. The whole
spectrum of the ground band can be obtained from the set
$\{E_{\gamma}\}$. Obviously, this equation is a generalization
of the one for spin-1 condensates given in ref.\cite{bao04} It
is also a more accurate version for the one given as Eq.(74) in
ref.\cite{mue}, in which the spin-dependent interaction has
been neglected.

\section{Thomas-Fermi approximation}

Since $N$ is usually large, Eq.(6) can be solved by using the
TFA. Neglecting the kinetic term in Eq.(6), when $\hbar\omega$
and $\lambda$ are used as units, the normalized TFA solution is
$\phi_{\gamma}=\sqrt{\frac{15}{8\pi r_0^3}}\sqrt{1-(r/r_0)^2}$
and $\epsilon_{\gamma}=r_0^2/2$, where
\begin{equation}
 r_0
  =  (\frac{15}{2\pi N}\tilde{E}_{vS})^{1/5},
 \label{7}
\end{equation}
is the TF-radius. From this solution we can obtain the root
mean square radius $R_{\mathrm{rms}}\equiv\langle
r^2\rangle^{1/2}=\sqrt{3/7}r_0$, and the total energy
\begin{equation}
 E_{\gamma}
  =  \frac{1}{7}
     ( \frac{3^2 5^7}{2^7\pi^2} )^{1/5}
     N^{3/5}
     ( \tilde{E}_{vS})^{2/5}
  =  \frac{5}{6}
     N
     R_{\mathrm{rms}}^2.
 \label{8}
\end{equation}
Thus, the total energy is directly related to the size
(measured by $R_{\mathrm{rms}}$).

When $N$ is large, we know from Eq.(4) that $\tilde{E}_{vS}$ is
proportional to $N^2$. Accordingly, $R_{\mathrm{rms}}\propto
N^{1/5}$ and $E_{\gamma}\propto N^{7/5}$. Thus the size
increases with $N$ very slowly, but $E_{\gamma}$ increases with
$N$ faster than linearly.

On the other hand, it is noted that, in the units of
$\hbar\omega$ and $\lambda$, all the strengths
$g_s\propto\sqrt{\tilde{\omega}}$, where $\tilde{\omega}$ is
the magnitude of $\omega$, i.e., $\omega=\tilde{\omega}s^{-1}$.
Therefore $\tilde{E}_{vS}\propto\sqrt{\tilde{\omega}}$. When
$\omega$-independent units are used, the size is described by
$R_{\mathrm{rms}}\lambda\propto(\tilde{\omega})^{-2/5}$, i.e.,
a stronger trap leads to a smaller size. Similarly, one can
prove that the chemical potential
$\epsilon_{\gamma}\hbar\omega\propto(\tilde{\omega})^{6/5}$,
and the total energy
$E_{\gamma}\hbar\omega\propto(\tilde{\omega})^{6/5}$. Thus the
total energy will increase with $\omega $\ a little faster than
linearly.

\section{Ground state and its phase}

It is obvious from Eq.(4) that which pair of $v$ and $S$ is
more advantageous to binding depends on the competition between
$g_v$ and $g_S$, which depend on $g_{04}$ and $g_{24}$, and
they are considered as variable. Therefore, we introduce a
variable parameter $\varphi$ to manifest the competition. In
the units of $\hbar\omega$ and $\lambda$, the strengths are
assumed as $g_{04}=u(\tilde{\omega})^{1/2}\cos\varphi$,
$g_{24}=u(\tilde{\omega})^{1/2}\sin \varphi$,
$g_{(024)}=2.5u(\tilde{\omega})^{1/2}$, and $u=10^{-3}$ (For a
comparison, the realistic strengths for $^{87}$Rb are
$g_{24}=-0.28u(\tilde{\omega})^{1/2}$,
$g_{04}=-0.41u(\tilde{\omega})^{1/2}$, and
$g_{(024)}=2.39u(\tilde{\omega})^{1/2}$). In fact, it has been
shown previously that the phase of the g.s. depends on
$\varphi$.\cite{cvc,mko,mue}

\begin{figure}[tbp]
 \centering \resizebox{0.95\columnwidth}{!}{\includegraphics{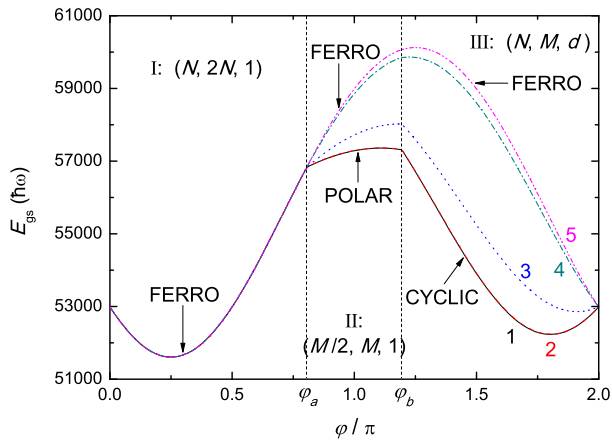} }
 \caption{(color on line) The ground state energy
$E_{\mathrm{gs}}$ (in $\hbar\omega$) versus $\varphi$.
$N=12000$ and $\omega=300s^{-1}$ are assumed (the same in the
following figures except specified). The total magnetization is
given at five cases as $M=0$, $1000$, $N$, $2N-1000$, and $2N$,
respectively, for the curves "1" to "5". There are three
regions of $\varphi$ separated by
$\varphi_a=\tan^{-1}(\frac{7(N+M/2+3)}{-10(N+M/2-2)})$ and
$\varphi_b=\tan^{-1}(\frac{-7}{-10})$ marked by the two
vertical dotted lines. In each region the g.s. has its specific
$(v,S,d)$ marked in the region, where $d$ is the degeneracy. In
region I, $E_{\mathrm{gs}}$ is $M$-independent. In I and II,
the g.s. is not degenerate ($d=1$). In III the degeneracy
depends on $M$ seriously. For the curves "1" to "5", $d=1$,
$167$, $2001$, $167$, and $1$, respectively.}
 \label{fig:1}
\end{figure}

Under the TFA, the ground state energy $E_{\mathrm{gs}}$ is
just the lowest $E_{\gamma}$ given by Eqs.(8) and (4). Although
$E_{\gamma}$ does not directly depend on $M$, the condition
$S\geq|M|$ restricts the possible choice of $S$, and
$E_{\mathrm{gs}}$ is thereby affected. $E_{\mathrm{gs}}$ versus
$\varphi$ with $M$ given at five values are shown in Fig.1.
There are two critical points
$\varphi_a=\tan^{-1}(\frac{7(N+M/2+3)}{-10(N+M/2-2)})$ and
$\varphi_b=\tan^{-1}(\frac{-7}{-10})$.\cite{pvi} When $\varphi$
goes through either $\varphi_a$ or $\varphi_b$ a phase
transition will occur. This phenomenon was found as early as in
2000 by using the mean field theory (in which
$\varphi_a=\tan^{-1}(\frac{7}{-10})$ and the same $\varphi_b$
were found.\cite{cvc} The present more accurate result will
tend to the old one when $N$ is large).

When $\varphi$ is in $(0,\varphi_a)$, i.e., in region-I, the
g.s. does not depend on $M$ and will have $(v,S,d)=(N,2N,1)$,
where $d$ is the degeneracy of the state. Since all the spins
align along a common direction in this g.s., it is in the
ferro-phase.\cite{cvc,mko,mue}

When $\varphi$ is in $(\varphi_a,\varphi_b)$, i.e., in
region-II, the g.s. depends on $M$ and will have
$(v,S,d)=(M/2,M,1)$. Since $v=M/2$ and $S=M$, all the $M/2$
unpaired spins must align along a common direction so that the
total spin can be maximized (i.e., $S$ can be equal $M$). Thus
each g.s. in region-II is a mixture of a group of aligned spins
together with the $(N-M/2)/2$ singlet pairs, namely, a mixture
of ferro-phase and polar phase. When $M=0$, $(v,S,d)=(0,0,1)$,
thus it is in pure polar phase (we have assumed that $N$ is
even, otherwise $S=2$). When $M$ is larger, more spins are in
the ferro-phase, and accordingly $E_{\mathrm{gs}}$ gets higher
as shown in the figure. When $M=2N$, $(v,S,d)=(N,2N,1)$, and
the g.s. is in pure ferro-phase.

When $\varphi$ is in $(\varphi_b,2\pi )$, i.e., in region-III,
the g.s. depends also on $M$ and has $(v,S,d)=(N,M,d)$. Note
that these g.s. have $v=N$, therefore they do not contain any
singlet pairs, but the spins in triplets are allowed. It is
recalled that, in regions I and II, the g.s. has $S=2v$. It
implies that all the unpaired spins must be aligned, and
therefore there is no room for the triplets. Whereas in III,
the number of triplets $n_{\mathrm{tri}}$ can have different
choices in a specific domain under the constraints of symmetry
given in points (i) to (iv) in the paragraph above Eq.(4)
(while those spins not in the triplets are coupled to $S=M$).
This leads to the degeneracy of the g.s.\cite{agh} When $M=0$
the g.s. has $S=0$ and therefore $N_{\mathrm{free}}=0$ (point
(ii)). Thus, all the particles are in the triplets (i.e., in a
pure cyclic phase), and henceforth $d=1$ (we have assumed that
$N$ is a multiple of 3 to simplify the discussion). When
$M=2N$, the g.s. has $S=2N$ and therefore
$N_{\mathrm{free}}=N$. Thus, neither the pairs nor the triplets
emerge (i.e., in a pure ferro-phase), and henceforth $d=1$
also. When $M$ is neither 0 nor $2N$, the g.s. is in general
degenerate. In particular, when $M=N$, $N_{\mathrm{free}}$ can
be ranged from $M/2$ to $M$ (each step is 3). In this case the
degeneracy $d$ is maximized (the curve "3" in III has
$d=2001$). The appearance of highly degenerate g.s. in the
region III is a notable feature.

It is reminded that the polar phase is composed of $s=0$ pairs.
Therefore, when $g_0$ is less positive than $g_2$ and $g_4$,
the g.s. will prefer this phase. The cyclic phase is composed
of $s=0$ triplets in which every pair of spins must be coupled
to $s=2$. Therefore, when $g_2$ is less positive, the g.s. will
prefer this phase. While the ferro-phase is composed of aligned
spins, in which every two must be coupled to $s=4$. Therefore,
when $g_4$ is less positive, the g.s. will prefer this phase.
Thus the phase transitions manifested above reflects the
competition of the interactions of the three $s$-channels. The
competition depends on $M$ because the least number of spins
that must be aligned is determined by $M$.

When $M=2N$ is given, the ferro-phase is the only choice
disregarding how the interaction is as shown by the uppermost
curve in Fig.1. When $N$ is sufficiently large the terms $\sim
1/N$ can be neglected, then we have
\begin{eqnarray}
 \tilde{E}_{vS}
 &\approx&
    ( a_2+g_v+4g_S)N^2 \nonumber \\
 &=& N^2
     [-\frac{1}{6}
       (\cos\varphi
       +\sin\varphi)
      +1.25]
     u(\tilde{\omega})^{1/2}.
 \label{9}
\end{eqnarray}
From this formula we know that $\tilde{E}_{vS}$ has a minimum
at $\varphi=\pi/4$ and a maximum at $\varphi=5\pi/4$ where
$\tilde{E}_{vS}=(\frac{15\sqrt{2}\mp 4}{12\sqrt{2}})N^2
u(\tilde{\omega})^{1/2}$. Since the total energy is
proportional to $N^{3/5}(\tilde{E}_{vS})^{2/5}$ as shown in
Eq.(8), with $N=12000$ and our parameters, the minimum at
$\varphi=\pi/4$ has $E_{\mathrm{gs}}=51612\hbar\omega$, and the
maximum at $\varphi=5\pi/4$ has
$E_{\mathrm{gs}}=60125\hbar\omega$. This explains the origin of
the dip and the highest peak in Fig.1.

Since the root mean square radius is an observable, it is
desirable to study $R_{\mathrm{rms}}$ so that the theory can be
checked via experimental measurement. According to Eq.(7),
$R_{\mathrm{rms}}$ is proportional to
$(E_{\mathrm{gs}})^{1/2}$, thus the variation of
$R_{\mathrm{rms}}$ versus $\varphi$ with the minimum and
maximum is predicted. In particular, when $\varphi=\pi/4$
($5\pi/4$) and the unit $\lambda$ is replaced by $\mu m$, the
minimum (maximum) in Fig.1 is associated with
$R_{\mathrm{rms}}=3.53$ (3.81) $\mu m$. In region II and III,
it is predicted that the increase of $M$ would lead to the
increase of the size. This is also a point to be checked.

\section{Stability of the ground state}

\begin{figure}[tbp]
 \centering \resizebox{0.95\columnwidth}{!}{\includegraphics{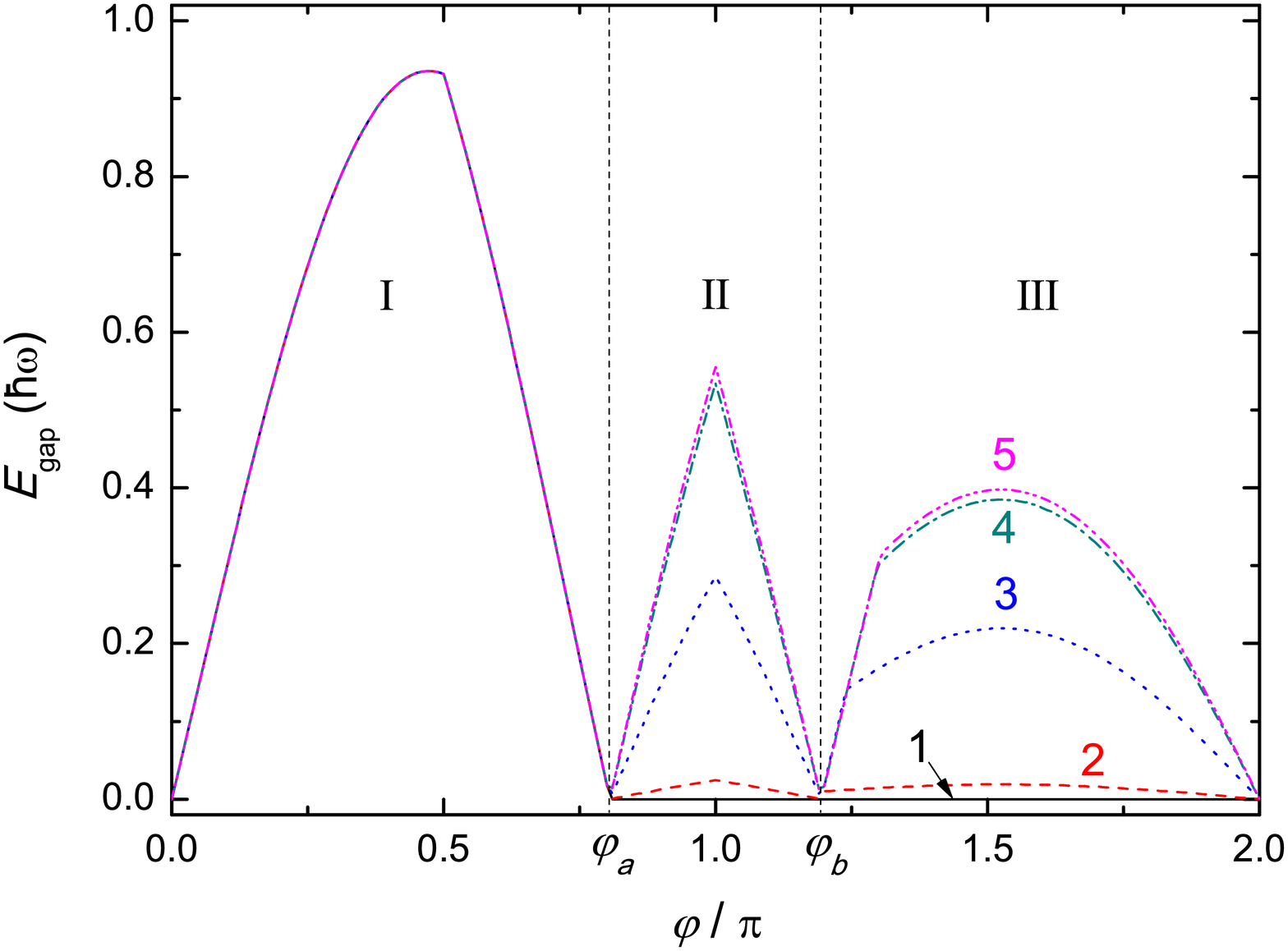} }
 \caption{(color on line) The energy gap $E_{\mathrm{gap}}$ (in
$\hbar\omega$) versus $\varphi$. The parameters are the same as
in Fig.1, except that the curve "5" has $M=2N-4$ instead of
$M=2N$.}
 \label{fig:2}
\end{figure}

It is believed that the stability of the g.s. is assured by the
gap $E_{\mathrm{gap}}$, namely, the energy difference between
the first excited state and the g.s.. The former has its
$(v,S)$ slightly different from that of the g.s., $(v_g,S_g)$,
and can be easily obtained from Eq.(4). An example of
$E_{\mathrm{gap}}$ is shown in Fig.2. Note that when $M=2N$,
$(v,S)=(N,2N)$ is the only choice, no choice other than
$(N,2N)$ is allowed. Thus, under $M$-conservation, there are no
excited spin-modes. Hence, in order to show the gap for a very
large $M$, the curve "5" in Fig.2 is not given as $M=2N$ but
$M=2N-4$. Fig.2 has the following features:

(i) Since Eq.(4) manifests that the energy is nearly $\propto
v^2$ and $S^2$, the gap would be in general large if both $v_g$
and $S_g$ are large. In this case, a small deviation in $(v,S)$
will cause a great change in energy. Whereas if one of them is
zero, the gap would be much lower. In region I the g.s. is in
the ferro-phase with $(v_g,S_g)=(N,2N)$. Therefore, the gap is
very high and accordingly \textit{the g.s. is very stable in
the ferro-phase}. In II and III, the upmost curve in Fig.2 has
$M=2N-4$. Accordingly, $(v_g,S_g)=(N-2,2N-4)$ and $(N,2N-4)$,
respectively. Since they are also large, the gap is also high.
When $M=2N-4$, the g.s. is very close to be in ferro-phase and
therefore is stable. However, when $M$ decreases, the stability
reduces. In particular, when $M=0$, $(v_g,S_g)=(0,0)$ and
$(N,0)$ in II and III, respectively. In this case
$E_{\mathrm{gap}}\leq 0.00012$ was found in both II and III as
shown by the lowest curve in Fig.2. Thus \textit{the g.s. in
polar and cyclic phases with a small} $M$ \textit{are highly
unstable}.

(ii) $E_{\mathrm{gap}}$ in every region appears as a peak,
namely, in the middle part it is higher but very low when
$\varphi$ is close to the borders. For Fig.2 the g.s. in I has
$(v,S)=(N,2N)$, while the first excited state has
$(v,S)=(N,2N-2)$ if $\varphi\leq 0.4722\pi$. Thus the
excitation from $(N,2N)$ to $(N,2N-2)$ is caused by a change in
$S$. However, when $\varphi\rightarrow 0$, accordingly
$g_S\rightarrow 0$. Therefore the gap is zero. On the other
hand, the first excited state has $(v,S)=(N-2,2N-4)$ when
($0.4722\pi<\varphi\leq\varphi_a$). From Eq.(4), we know the
gap
\begin{eqnarray}
 &&\tilde{E}_{N,2N}-\tilde{E}_{N-2,2N-4} \nonumber \\
 &=& \frac{4}{70}
     [ 7(N+\frac{1}{2})g_{04}+10(N-2)g_{24}].
 \label{10}
\end{eqnarray}
One can prove that\ the gap is zero when
$\varphi=\varphi_{a'}\equiv\tan
^{-1}\frac{7(N+1/2)}{-10(N-2)}$. Since $\varphi _{a^{\prime }}$
is extremely close to $\varphi _{a}$, this explains the decline
of the gap when $\varphi \rightarrow \varphi _{a}$. Based on
Eq.(4), the decline of the peak in II and III can be similarly
explained. Thus, \textit{in the neighborhoods of the borders,
the g.s. is highly unstable.}

In addition to the gap, another factor that could affect the
stability of the g.s. is the level density in the neighborhood
of the lowest level. This density can be calculated based on
Eq.(8). As an example, the number of levels with excitation
energy $\leq 0.1\hbar\omega$ is given in Table \ref{tab:1}, in
which seven choices of $\varphi$ and four choices of $M$ are
chosen. The other parameters are the same as in Fig.1. If the
degeneracy of a level is $d$, then the number contributed by
this level is $d$.

\begin{table}[htb]
  \caption{The number of levels lower than $0.1\hbar\omega$
versus $M$ and $\varphi$. $\varphi_a$ and $\varphi_b$ denote
the borders, $\delta=0.01\pi$. The first column gives the
region of $\varphi$.}
 \label{tab:1}
  \begin{center}
    \begin{tabular}{c|c|c|c|c}
      \hline\hline
       Region   &   $\varphi$           &   \multicolumn{3}{c}{Number of levels}    \\      \cline{3-5}
                &                       &   $M=0$       &   $M=N$   &   $M=2N-4$    \\
      \hline
       I        &   $\pi/4$             &   $1$         &   $1$     &   $1$         \\
       I        &   $\varphi_a-\delta$  &   $4$         &   $4$     &   $2$         \\
       II       &   $\varphi_a+\delta$  &   $23059$     &   $7$     &   $2$         \\
       II       &   $\pi$               &   $20596$     &   $1$     &   $1$         \\
       II       &   $\varphi_b-\delta$  &   $167171$    &   $19$    &   $2$         \\
       III      &   $\varphi_b+\delta$  &   $3372$      &   $7999$  &   $2$         \\
       III      &   $3\pi/2$            &   $885$       &   $2001$  &   $1$         \\
      \hline\hline
    \end{tabular}
  \end{center}
\end{table}

From the table we know that

(i) When $\varphi$ is in region I or $M$ is close to $2N$
(i.e., in or close to the ferro-phase), the density of
low-lying states is rather diffuse. Together with the big gap,
both factors assure the stability of the g.s. (When the number
$=1$ as shown in the second row, there is no excited states
lower than $0.1\hbar\omega$).

(ii) When $\varphi $ is in region II and $M$ is very small
(i.e., in or close to the polar phase), the density is
extremely dense. Together with the very small gap, both factors
lead to a highly unstable g.s.. This situation can be greatly
improved when $M$ gets larger.

(iii) When $\varphi$ is in region III and $M$ is not very large
(say, $M\leq N$), the low-lying spectrum is also dense but not
as dense as in the polar phase. Note that, when $M$ is not
close to 0 or $2N$ (say, $0.05\times 2N\leq$ $M\leq 0.95\times
2N$), it has been mentioned that the states in cyclic phase
could be highly degenerate. The great degeneracy contributes to
the level-density substantially (say, in the last row of Table
\ref{tab:1}, the number 2001 arises completely from the
degeneracy of the g.s.).

\section{Numerical solutions of the symmetry adapted GP-equation}

By making use of the TFA we have obtained analytical solutions,
thereby the related physics can be understood in an analytical
way. In order to evaluate the accuracy of the TFA, the equation
with the kinetic energy included is solved numerically and the
results are given below for a comparison. Firstly, we found
that the curves for $E_{\mathrm{gs}}$ from the numerical
solutions as a whole is an upward shift of those plotted in
Fig.1. It implies that the amount of total kinetic energies
contained in various states with very different spin-textiles
are similar, and the $E_{\mathrm{gs}}$ given under the TFA are
correct when $N=12000$ (except the omission of the kinetic
energy). The shift is about $3.5\times 10^{3}\hbar\omega$ (thus
the kinetic energy is about $6\%$ of the total energy in our
cases). Secondly, we found that the curves for
$E_{\mathrm{gap}}$ from the numerical solutions overlap the
curves from TFA plotted in Fig.2. It implies that the amount of
kinetic energies contained in the g.s. and in the first excited
state is similar. Thus the spin-excitation does not remarkably
affect the spatial motion.

\begin{widetext}
\begin{table}[htb]
 \caption{Root Means Square Radius $R_{\mathrm{rms}}$ in $\mu m$
from numerical calculation versus $N$, $\varphi$, and $M$ with
$\omega=300s^{-1}$. The values from TFA are given inside the
parentheses.}
 \label{tab:2}
  \begin{center}
    \begin{tabular}{c|c|c|c||c|c|c|c||c|c|c|c}
      \hline\hline
       $N$  &   $\varphi$   &   $M$         &  $R_{\mathrm{rms}}$   &   $N$     &   $\varphi$   &   $M$         &   $R_{\mathrm{rms}}$  &   $N$     &   $\varphi$   &   $M$         &   $R_{\mathrm{rms}}$  \\
      \hline
       1200 &   $\pi/4$     &   $0,N,2N$    &  $2.48(2.23)$         &   6000    &   $\pi/4$     &   $0,N,2N$    &   $3.19(3.07)$        &   12000   &   $\pi/4$     &   $0,N,2N$    &   $3.62(3.53)$        \\      \cline{2-4} \cline{6-8} \cline{10-12}
            &   $\pi$       &   $0$         &  $2.57(2.35)$         &           &   $\pi$       &   $0$         &   $3.34(3.24)$        &           &   $\pi$       &   $0$         &   $3.80(3.72)$        \\      \cline{3-4} \cline{7-8} \cline{11-12}
            &               &   $N$         &  $2.58(2.36)$         &           &               &   $N$         &   $3.36(3.25)$        &           &               &   $N$         &   $3.81(3.73)$        \\      \cline{3-4} \cline{7-8} \cline{11-12}
            &               &   $2N$        &  $2.60(2.38)$         &           &               &   $2N$        &   $3.39(3.29)$        &           &               &   $2N$        &   $3.85(3.77)$        \\      \cline{2-4} \cline{6-8} \cline{10-12}
            &   $5\pi/4$    &   $0$         &  $2.56(2.33)$         &           &   $5\pi/4$    &   $0$         &   $3.33(3.22)$        &           &   $5\pi/4$    &   $0$         &   $3.78(3.70)$        \\      \cline{3-4} \cline{7-8} \cline{11-12}
            &               &   $N$         &  $2.58(2.35)$         &           &               &   $N$         &   $3.35(3.25)$        &           &               &   $N$         &   $3.80(3.73)$        \\      \cline{3-4} \cline{7-8} \cline{11-12}
            &               &   $2N$        &  $2.62(2.40)$         &           &               &   $2N$        &   $3.42(3.32)$        &           &               &   $2N$        &   $3.88(3.81)$        \\
      \hline\hline
    \end{tabular}
  \end{center}
\end{table}
\end{widetext}

Furthermore, the size of the system is expected to increase by
including the kinetic energy. The results from the numerical
calculation are shown in Table \ref{tab:2}, where the
$R_{\mathrm{rms}}$ from TFA are given inside the parentheses
and are smaller. From the table we know that the TFA leads to a
$\sim 8\%$ reduction of the $R_{\mathrm{rms}}$ (if $N=1200$),
or a $\sim 2\%$ reduction of the $R_{\mathrm{rms}}$ (if
$N=12000$). Nonetheless, the two features found previously
under the TFA remain unchanged, namely, (i) When the g.s. is
not in the ferro-phase, a larger $M$ leads to a larger size.
(ii) When the g.s. is in the ferro-phase
($\varphi\leq\varphi_a$ or $M=2N$), the size is minimized when
$\varphi=\pi/4$ and is maximized when $\varphi=5\pi/4$.

\begin{figure}[tbp]
 \centering \resizebox{0.95\columnwidth}{!}{\includegraphics{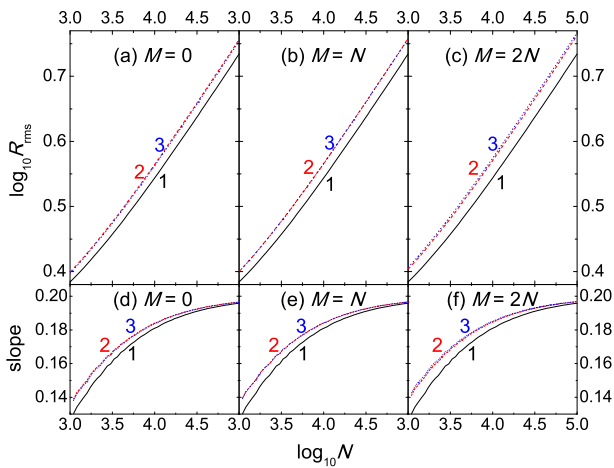} }
 \caption{(color on line) $\log_{10}R_{\mathrm{rms}}$ versus
$\log_{10}N$ with $\omega=300s^{-1}$ where $R_{\mathrm{rms}}$
is in $\mu m$. $M$ is given at three values. $\varphi=\pi/4$,
$\pi$, and $5\pi/4$ for the curves "1" to "3", respectively.
The lower panels are for the slopes of the curves.}
 \label{fig:3}
\end{figure}

\begin{figure}[tbp]
 \centering \resizebox{0.95\columnwidth}{!}{\includegraphics{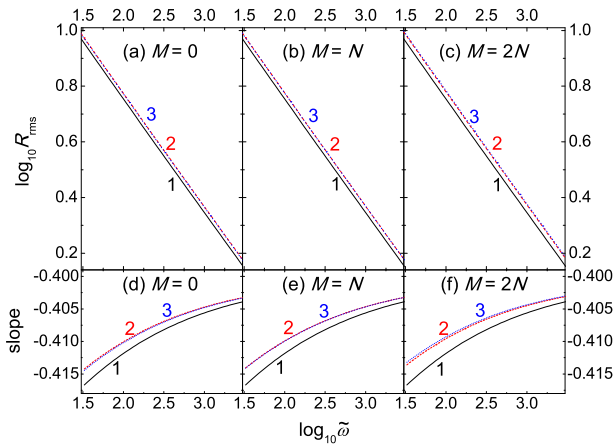} }
 \caption{(color on line) $\log_{10}R_{\mathrm{rms}}$ versus
$\log_{10}\tilde{\omega}$ with $N=12000$. Refer to Fig.3. Note
that the vertical scale in 4d to 4f is very small. It implies
that the curves in 4a to 4c are very close to straight lines.}
 \label{fig:4}
\end{figure}

It has been predicted based on the TFA that
$R_{\mathrm{rms}}\propto N^{1/5}$. To check this relation
$\log_{10}R_{\mathrm{rms}}$ from the numerical solutions are
plotted in Fig.3 versus $\log_{10}N$. In 3a to 3c the curves
are not exactly straight lines. Their slopes depend on $N$ and
are plotted in 3d to 3f, respectively. Disregarding $M$ and
$\varphi$ the slopes tend to $1/5$ when $N$ increases as
predicted. It has been predicted based on the TFA that
$R_{\mathrm{rms}}\lambda \propto(\tilde{\omega })^{-2/5}$. To
check this relation $\log_{10}R_{\mathrm{rms}}$ from the
numerical solutions are plotted in Fig.4 versus
$\log_{10}\tilde{\omega}$. The slopes of the curves tend to
$-2/5$ when $\tilde{\omega}$ increases as predicted.

\section{Summary}

The generalized GP equation adapted to the $U(5)\supset
SO(5)\supset SO(3)$ symmetry has been derived for spin-2
condensates. This equation has been solved analytically under
the TFA and by strict numerical calculation. It was found that
the TFA is applicable if $N$ is large (say, $N\geq 10^{4}$).
The emphasis is placed on the g.s.. Based on a rigorous
treatment of the spin-degrees of freedom, the detailed
spin-textiles, i.e., the ferro-, polar, and cyclic phases, and
their mixing, are explained in a many-body way and thus the
underlying physics can be understood beyond the
mean-field-theory. Besides, the variation of the spin-textiles
versus $M$ in regions II and III is notable. Note that the
factors affecting the stability of the g.s., such as the gap
and the neighboring level density, together with the degeneracy
of the g.s. itself, are less touched in existing literatures.
These factors are studied in detail in this paper. The great
difference in the stability and degeneracy caused by varying
$\varphi$ (which marks the features of the interaction) and $M$
is notable (this is true even when $\varphi$ varies within a
region, i.e., the g.s. remains in the same phase). We believe
that the effect of these factors would be serious when the
temperature is very low. Since $R_{\mathrm{rms}}$ is an
observable, efforts have also been made to clarify the relation
between $R_{rms}$ and $N$, $\omega$, and $\varphi$. This
provides a way for checking the theories with experimental
data.

\begin{acknowledgments}
The project is supported by the National Basic Research Program
of China under the grant 10874249.
\end{acknowledgments}


\begin{thebibliography}{9}
\bibitem{a1}
 When $N\rightarrow\infty$, $\theta_{\mathrm{tri}}$ tends to
$(\chi\chi\chi)_0$. Otherwise, a revision depending on $N^{-1}$
is necessary. The details are referred to \cite{pvi,ech}.
\end{thebibliography}
\end{document}